\documentstyle[preprint,aps,eqsecnum]{revtex}
 
\begin{document}
\def\u{{\cal U}}
\def\v{{\cal V}}
\def\x{{\cal X}}
\def\t{{\cal T}}
\def\be{\begin{equation}}
\def\ee{\end{equation}}
\draft
\preprint{Alberta-Thy-26-94}

\title{The  Affine symmetry of self-dual gravity}
\author{Viqar Husain\footnote{Address after September 15. '95:
Center for Gravitational Physics and Geometry, The Pennsylvania
State University, University Park, PA 16802-6300, USA. }}

\address{Dept. of Mathematics and Statistics, University of Calgary,\\
Calgary, Alberta, Canada T2N 1N4.}
 \maketitle

\begin{abstract}
Self-dual gravity may be reformulated as  the  two dimensional
principal chiral  model with the group of area preserving diffeomorphisms
as its gauge group.  Using this formulation,  it is shown that
self-dual gravity contains an  infinite dimensional hidden symmetry
whose generators form the Affine (Kac-Moody)  algebra associated with
the Lie algebra  of area preserving diffeomorphisms.  This result
provides an observable algebra and a solution generating technique
for self-dual gravity.
\end{abstract}
\bigskip
\pacs{PACS numbers: 02.20.Tw, 04.90.+e, 11.30.-j}

\vfill
\eject

\section{Introduction}

Geometric field theories are of interest as models for quantum gravity.
 Topological field theories are examples of such theories  that
have a finite number of degrees of freedom. As such they are
usually completely  solvable quantum mechanically, and may
be compared, in this respect,  with the minisuperspace reductions of
general relativity.

It is of more interest however to look for field theoretic models with
an infinite number of degrees of freedom for modelling quantum gravity,
since these  are more likely to capture the main features of
dynamical geometries.   The types of models that have been studied
include midisuperspace reductions of general relativity and self-dual
gravity.

For the quantization of  model theories, one approach involves finding
 a suitable Poisson algebra of observables at the classical level, which
is to be represented as an operator algebra in the quantum theory.  The
problem of  finding classical  observables that form a nice algebra is not
an easy task  for a field theory, and most of the examples where this can
be done are the integrable two dimensional theories, such as the KdV and
Sine-Gordon equations, which are not geometric field theories
in the sense required for quantum gravity.

In the cases where a suitable classical Poisson algebra can be found,
quantization involves finding a representation of this algebra such that
the observables are represented as Hermitian operators.  This also may
not be an easy task, especially  if the Poisson algebra  is infinite
dimensional, which is expected to be the case for a field theory.

In this paper we study self-dual gravity as a possible model for
quantum gravity.  We will  only  consider in detail some classical
aspects of the theory,  and will outline an approach to quantization.

 Self-dual gravity is defined by the statement that the
Riemann curvature $R$ of a metric $g$ is proportional to its dual
$^\star R$ :
\be
  ^\star R_{abcd} := {1\over 2} \epsilon_{ab}^{\ \ ef}\ R_{efcd}
= \lambda\ R_{abcd}\ ,
\ee
where $\lambda$ is the proportionality factor and $\epsilon$ is
the totally antisymmetric tensor density. $\lambda$ is determined
by noting from the definition of duality that
$^{\star\star}R = {\rm det}(g^{ab})\ R$.
Therefore, taking the dual of  $^{\star}R=\lambda\ R$ gives
\be
^{\star\star}R = \lambda^\star\  R = \lambda^2\ R.
\ee
Hence $\lambda = \pm \sqrt{{ \rm det}(g^{ab})}$.  This shows that
self-dual gravity has non-flat solutions either for real metrics
of signature $(--++)$ or $(++++)$, or  for complex metrics.

Since the Riemann curvature  may be split into its self-dual and
anti-self-dual parts, imposing  self-duality  gives a reduction
of general relativity with only one local  degree of freedom.
Self-duality implies the vanishing of the Ricci tensor
via the cyclic Bianchi identity $R_{a[bcd]}=0$ :
\begin{equation}
 R_{ac} = R_{abcd}\ g^{bd} = {1\over 2\lambda}\
\epsilon_{ab}^{\ \ ef}\ R_{efc}^{\ \ \ b} = 0.
\end{equation}
 This reduction therefore provides a non-trivial diffeomorphism invariant
geometric field theory which may be studied classically and quantum
mechanically.

An action principle, or equivalently a Hamiltonian formulation for
self-dual gravity is not known, although the approach presented
in this paper suggests a way of obtaining this.

Most work on this theory has been classical, and it is believed to
be integrable. The main result that indicates integrability is
Penrose's twistor construction of the general solution \cite{pen}.
There are a number of other interesting results associated with
this theory: There is a formulation  using the Ashtekar
Hamiltonian variables which give the self-duality equations in first
order form as the evolution equations of  triad of vector fields
\cite{ajs};
the latter form of the equations may be obtained from a
0+1 reduction of the self-dual Yang-Mills equation \cite{masnew};
the Plebanski equation \cite{pleb}, which is another formulation
of the self-duality equation, may be  derived  from the  two
dimensional Wess-Zumino equation \cite{park};  this latter work also
identifies the infinite dimensional area preserving diffeomorphism
 algebra as a special symmetry associated with the theory.

Recently another formulation of self-dual gravity has been given by
the author \cite{vh1}, which shows how this theory may be
rewritten as the principal chiral model with the group of area preserving
diffeomorphisms as its gauge group. Similar results have  also been
obtained by Ward \cite{ward}.  This two dimensional formulation
allows the use of a  method \cite{bizz} for obtaining an infinite number
of  conserved currents  for self-dual gravity \cite{vh2}. It thus
provides another hint that the theory is integrable, since the principal
chiral model with gauge group $SU(N)$ is known to be integrable
\cite{fadtakh}.  It is likely that the two-dimensionsal techniques
used to prove integrability will also be applicable to the
infinite dimensional Lie algebra case relevant for self-dual
gravity.

The chiral model formulation also provides an approach to
obtaining a Hamiltonian formulation for self-dual gravity,
since the Hamiltonian formulation of the chiral model for $SU(N)$ is
known.  However, again  the generalization of  this to the case of the
infinite dimensional Lie algebra associated with self-dual gravity
requires further study.

In this paper we will show that the conserved currents of self-dual
gravity generate an infinite dimensional symmetry on the solution space
of  the theory, and that its generators form an  Affine  algebra
associated with the Lie algebra  of area preserving diffeomorphisms.
This result identifies the observable algebra of self-dual
gravity,  since  it is expected that symmetries on the solution
space  correspond to symmetries on the phase space generated
via Poisson brackets.

( Infinite dimensional  symmetry transformations on the solution space
of two Killing field reduced
general  relativity were found by Geroch \cite{ger}, and later shown
to form an SL(2,R) Affine algebra \cite{wu}.  Recently,
the phase space realization of the solution space generators of the
Geroch transformations have been found by the author \cite{vh3}.)

In the next Section  we review the chiral model formulation of self-dual
gravity. In Section III   the symmetry transformations  on the solution
space are identified and their  algebra is calculated.  Section IV
contains a  summary  and discussion of  a  possible approach to
quantization.

\section{Self dual gravity as the principal chiral model}

 Self dual gravity may be reformulated as the chiral model in two dimensions
\cite{vh1}. This may be seen by starting with the Ashtekar-Jacobson-Smolin
(AJS) form of the self-duality equations, which were derived from the
Ashtekar  Hamiltonian formulation of general relativity
in Ref. \cite{ajs}.  The AJS equations are  evolution
equations for three divergence free vector fields $V_i^a(t,x,y,z)$  given
on a  three dimensional  surface with a  fixed volume element, which may
be written as  $dx \wedge dy\wedge dz$ in local coordinates.
The indices $a,b...$ are four dimensional, but in the following they
will also be used on vectors that are tangent to three and two
dimensional surfaces. The indices $i,j...=1,2,3$ label the vector field.
The AJS self-duality equations are
\be   \partial_a V^a_i=0, \label{div} \ee
\be  {\partial V^a_i \over \partial t} =
{1\over 2}\ \epsilon^{ijk}\ [\ V_j,V_k\ ]^a.
\label{evol} \ee
where $[\  , ]$ is the Lie bracket, and the divergence in (\ref{div}) is
 defined  with respect to the fixed volume element on the three-surface.
The self-dual metric is constructed from a  solution of these equations
using
\be
g^{ab} = ({\rm det} V)^{-1}\ [\ V^a_i V^b_j \delta^{ij} + V_0^a V_0^b\ ],
\ee
where we have defined $V_0^a = ( \partial/\partial t)^a$.
Eqn. (\ref{evol}) may be written in a covariant form as
\be [\ V_0,V_i\ ]^a = \epsilon^{ijk} [\ V_j,V_k\ ]^a. \label{cevol}\ee

The principal chiral model is the two dimensional field theory whose
Lagrangian is
\be
L = {1\over 2}\ {\rm Tr}\ \bigl(\  \partial_\mu g^{-1}
\partial_\nu g\ \bigr)
\ \eta^{\mu\nu},
\ee
where $g(x,t)$ is a matrix group element and $\eta^{\mu\nu}$
$(\mu,\nu =0,1=t,x)$ is a fixed flat background metric with signature
$(++)$.  The equations of motion are
\be
\eta^{\mu\nu} \partial_\mu\ (\ g^{-1}\partial_\nu g\ )=0.
\ee
This equation may be written as two  first order equations by using
the Lie  algebra  valued 1-form $A_\mu := g^{-1}\partial_\mu g$ :
\be
\eta^{\mu\nu} \partial_\mu A_\nu =0 \label{con},
\ee
\be
F_{\mu\nu} = \partial_\mu A_\nu - \partial_\nu A_\mu
+ [\ A_\mu,\ A_\nu\ ] = 0. \label{cur}
\ee

To see how Eqns. (\ref{cevol}) may be rewritten as the chiral
model equations (\ref{con}-\ref{cur}) we first  rewrite the former
using the linear combinations
\begin{eqnarray}
\t &=& V_0+iV_1 \ \ \ \ \ \ \ \ \ \u= V_0-iV_1 \nonumber \\
\x &=& V_2-iV_3 \ \ \ \ \ \ \ \ \ \v=V_2+iV_3.
\end{eqnarray}
Then    equations  (\ref{cevol}) become
\be
[\ \t,\x\ ] = [\ \u,\v\ ]=0 \label{c1},
\ee
\be
 [\ \t,\u\  ] + [\ \x,\v\ ]=0 \label{c2}.
\ee
Using the coordinate freedom, we choose coordinates $t,x$ such that
\be
\t^a = ({\partial\over \partial t})^a \ \ \ \ \ \ \ \ \ \
\x^a = ({\partial\over \partial x})^a,
\ee
 with $\u$ and $\v$ arbitrary  except that they satisfy the  divergence
free  condition (\ref{div}) with respect to the volume form
$\omega =  dx \wedge dy \wedge dz$ defined by the local coordinates.
Since $\t^a$ and $\x^a$ commute, they are surface forming, and as we will
see below, this surface will be the chiral model background.

The final step in obtaining the chiral model equations for self dual
gravity is to make a specific choice for $\u$ and $\v$ without  losing
generality, such that  these vector fields are divergence free (\ref{div}),
and lead to no reduction in  the number of  local degrees of freedom.
 A choice satisfying these conditions is
\begin{eqnarray}
\u^a &=& ({\partial \over \partial t})^a + \alpha^{ba}\ \partial_b A_0
\label{u}\\
\v^a &=& ({\partial \over \partial x})^a + \alpha^{ba}\ \partial_b A_1\ ,
\label{v}
 \end{eqnarray}
where
\begin{equation}
\alpha^{ab} = ({\partial\over \partial y})^{[\ a}\otimes
 ({\partial\over \partial z})^{b\ ]} =
\left( \begin{array}{cc} 0 & 1 \\ -1 & 0 \end{array} \right)
\end{equation}
 is the antisymmetric tensor that is the
 inverse of the two form $(dy\wedge dz)_{ab}$ in the $y-z$ surface,
and $A_0(t,x;y,z)$ and $A_1(t,x;y,z)$ are two arbitrary functions.
(A proof that there is no loss of generality in choosing
$\u$ and $\v$ as in (\ref{u}-\ref{v}) is given in Ref. \cite{vh2}. )
Substituting these into (\ref{c1}-\ref{c2}) gives
 \begin{eqnarray}
 \alpha^{ab}\partial_b\  \bigl[\ \partial_0 A_1 - \partial_1 A_0 +
 \{A_0,A_1\}\ \bigr] &=& 0 \label{ch2}\\
 \alpha^{ab}\partial_b\ \big[\ \partial_0A_0 + \partial_1 A_1\
\bigr] = 0.
\label{ch1}
\end{eqnarray}
 where the bracket on the left hand side of equation (\ref{ch2}) is the
 Poisson bracket with respect to $\alpha^{ab}$,
 \begin{equation}
 \{A_0,A_1\} := \alpha^{ab}\partial_aA_0\partial_bA_1=
 \partial_y A_0 \partial_z A_1
 -\partial_z A_0 \partial_y A_1, \label{pb}
 \end{equation}
 and $\partial_0,\partial_1$ denote partial derivatives with respect to
 $t,x$ etc.
Equations (\ref{ch2}-\ref{ch1}) imply that the terms in their
 square brackets are equal to two arbitrary functions of $t$ and $x$,
which we write as
 \begin{eqnarray}
  \partial_0 A_1 - \partial_1 A_0 +
 \{A_0,A_1\} &=& \partial_0 F(t,x) +\partial_1 G(t,x) \label{chi2}\\
 \partial_0A_0 + \partial_1 A_1 &=& \partial_1 F(t,x) - \partial_0 G(t,x),
\label{chi1}
\end{eqnarray}
where $F,G$ are arbitrary functions of $t,x$.
With the redefinitions
\begin{equation}
a_0(t,x;y,z):= A_0 + G \ \ \ \ \ \ a_1(t,x;y,z) := A_1 - F,
\label{redef}
\end{equation}
(\ref{chi2}-\ref{chi1}) become
\begin{eqnarray}
f_{01}:= \partial_0 a_1 - \partial_1 a_0 +
 \{a_0,a_1\} &=& 0, \label{chir2} \\
 \partial_0 a_0 + \partial_1 a_1 &=& 0 \label{chir1}.
 \end{eqnarray}
 These are precisely the chiral model equations (\ref{con}-\ref{cur})
on the $x,t$ `spacetime', with $y,z$ treated as coordinates on an
`internal'  space. The commutator in (\ref{cur}) has been replaced by
the Poisson bracket (\ref{pb}). The gauge group
 is therefore the group of diffeomorphisms that preserve $\alpha^{ab}$ on
 the internal space,  which is the group of area preserving
diffeomorphisms.  (We note that the redefinitions (\ref{redef}) do not
alter the vector fields $\u$ and $\v$ in Eqns. (\ref{u}-\ref{v}) ).

\section{Conserved currents and the symmetry algebra}

The chiral model for the $SU(N)$ groups, where the zero curvature
equation (\ref{cur}) contains a matrix commutator, is known to have
an infinite dimensional hidden symmetry \cite{bizz,dolan}.  Here we
will show  explicitly that when the matrix commutator is replaced by
a Poisson bracket as in Eqn. (\ref{chir2}), the hidden non-local
symmetry  transformations  form an Affine algebra  associated with
the Lie algebra of area preserving diffeomorphisms.

We first define the covariant derivative acting on functions
$\Lambda(t,x;y,z)$:
\be
D_\mu \Lambda := \partial_\mu \Lambda + \{a_\mu, \Lambda\}.
\ee
where $\{\  ,\  \}$ is as defined in (\ref{pb}) and $\mu,\ \nu=0,1=t,x$
are the two dimensional `spacetime' indices.   Under the
transformation
\be
\delta_\Lambda a_\mu := D_\mu \Lambda\ , \label{trans}
\ee
the left hand sides of Eqns. (\ref{chir2}-\ref{chir1}) change to
\be
\delta_\Lambda (\partial_\mu a_\mu) = \partial_\mu (D_\mu \Lambda)
 \label{tran1}
\ee
\be
  \delta_\Lambda (f_{01}) = \{f_{01},\Lambda\}.
\label{tran2}
\ee
Therefore, under the transformation (\ref{trans}),
(\ref{chir2}) is invariant because $f_{01}=0$, and (\ref{chir1}) will
be  invariant  only if  $D_\mu \Lambda$ is  a conserved current.
 We will now show following  Refs. \cite{bizz,dolan}, that a
heirarchy of functions $\Lambda^{(n)}(t,x;y,z)$ may indeed be constructed
such that
\be
J_\mu^{(n)} := D_\mu \Lambda^{(n)} \label{ncur}
\ee
are conserved currents.

We first note that for an arbitrary function $\Lambda^{(0)} (y,z)$,
the current
\be
J_\mu^{(0)} := D_\mu \Lambda^{(0)} = \{a_\mu, \Lambda^{(0)} \}
\label{cur0}
\ee
is conserved by virtue of (\ref{chir1}).  Therefore there exists a
function $\Lambda^{(1)}(t,x;y,z)$ such that
 \be
J_\mu^{(0)} = \epsilon_{\mu\nu}\partial_\nu \Lambda^{(1)},
\ee
where $\epsilon_{\mu\nu}$ is the antisymmetric tensor.
We also have by Eqn. (\ref{chir1})   that
\be
a_\mu = \epsilon_{\mu\nu} \partial_\nu \phi
\label{amu}\ee
 for some function
$\phi(t,x;y,z)$.  Together  these give
\be
\Lambda^{(1)} = \int_{-\infty}^x dx' \ D_0 \Lambda^{(0)} =
\int_{-\infty}^x dx' \  \{a_0, \Lambda^{(0)} \}
= \{\phi, \Lambda^{(0)}\}.
\label{lam1}
\ee
We now define a second current by
\be
J_\mu^{(1)} : =D_\mu \Lambda^{(1)}.
\ee
It is straight forward to see that this is also conserved :
\begin{eqnarray}
\partial_\mu J_\mu^{(1)} &=& \partial_\mu (
 \{ \partial_\mu  \phi, \Lambda^{(0)}\}  +
\{a_\mu,  \{   \phi, \Lambda^{(0)}\} \}
) \nonumber \\
 &=& \partial_\mu
\{\epsilon_{\mu\nu} a_\nu, \Lambda^{(0)}\}  +
\{a_\mu,  \{ \epsilon_{\mu\nu} a_\nu, \Lambda^{(0)}\} \}
 \nonumber \\
&=& \{ f_{01}, \Lambda^{(0)} \} =0,
\end{eqnarray}
since $\partial_\mu \Lambda^{(0)} = 0$ and $f_{01}=0$.

Assuming now that the $n$th. current (\ref{ncur}) is conserved
implies that there is a function $\Lambda^{(n+1)}(t,x;y,z)$ such that
\be
J_\mu^{(n)} = \epsilon_{\mu\nu} \partial_\nu \Lambda^{(n+1)}.
 \label{lamn1}\ee
Using this to show that $J_\mu^{(n+1)} := D_\mu \Lambda^{(n+1)}$
is conserved will complete the induction:
\begin{eqnarray}
\partial_\mu J_\mu^{(n+1)} &=& D_\mu \partial_\mu \Lambda^{(n+1)}
= D_\mu \epsilon^{\mu\nu} J_\nu^{(n)}  \nonumber \\
&=& D_\mu \epsilon^{\mu\nu} D_\nu \Lambda^{(n)}
= \{ f_{01}, \Lambda^{(n)} \} =0.
\end{eqnarray}
Equations (\ref{ncur}) and (\ref{lamn1}) give the relation between
the successive  $ \Lambda^{(n)}$:
\be
\Lambda^{(n+1)}(t,x) = \int_{-\infty}^x dx'\  D_0 \Lambda^{(n)}(t,x')
\label{lamrec}
\ee

The conserved currents generated by this procedure are non-trivial
by construction. For  if we are given a solution where the $a_\mu$ are
not zero,  then the method will give non-vanishing currents
which are independent.  This is because the $(n+1)$th. current involves
an extra integral which makes it more non-local than the $n$th. one.

We would now like to see what the  algebra of  these currents is.
The  generators of the symmetry transformations (\ref{trans})  are
\be
T^{(n)}_\Lambda :=  \int dtdx\  (\delta^{(n)}_\Lambda a_\mu)
\ {\delta \over \delta a_\mu}
 = \int d^2x\  ( D_\mu \Lambda^{(n)})\ {\delta \over \delta a_\mu},
\label{gen}
\ee
and they act on the space of solutions of (\ref{chir2}-\ref{chir1})
via
\be
\delta^{(n)}_\Lambda  a_\mu = D_\mu \Lambda^{(n)}.
\label{solcom}
\ee
(Note that the generators $T^{(n)}_\Lambda$ defined here are
functions  of the `internal' coordinates $(y,z)$. They could equally
well be defined with additional integrals over $y$ and $z$, which
would follow the standard textbook definitions of generators. All the
results presented here go through with either definition because the
$y,z$ internal space integrals are `bystanders' in the calculations
below.)

  We recall that $a_\mu$ and  $\Lambda^{(n)}$ are functions
of {\it all} the coordinates $t,x,y,z$ for $n>0$.  These functions
may be expanded in a suitable orthonormal basis of functions
$f_\alpha(y,z)$ on the  `internal'  $y,z$ space as
\be
\Lambda^{(n)}(t,x;y,z) =
\sum_{\alpha}   f_\alpha(y,z)\ \Lambda^{(n)}_\alpha (t,x),
\ee
 where
\be
 \Lambda^{(n)}_\alpha (t,x) =
\int dydz\ f_\alpha(y,z)\ \Lambda^{(n)}(t,x;y,z).
 \label{basis}
\ee
and $\alpha$ labels the basis.  Such basis functions will satisfy a
closed  Poisson algebra
\be \{ f_\alpha, f_\beta \} = C_{\alpha\beta}^{\ \ \gamma}\ f_\gamma
\ee
where $ C_{\alpha\beta}^{\ \ \gamma}$ are the structure constants
of the area preserving diffeomorphism group  of the $y,z$ surface.
For example, if the internal space is a
torus,  $y$ and $z$ are angles, and the natural basis are the functions
$f_{mn}(y,z) = {\rm exp}[\ i(my+nz)\ ]$, with $\alpha \equiv (m,n)$.
Their Poisson algebra is
\be
\{ f_{m,n}, f_{p,q} \}  = (pn-mq)\ f_{m+p,n+q},
\ee
which identifies the structure constants.  Natural bases for the area
preserving diffeomorphisms  for a number of  surfaces have been
studied in Ref. \cite{winf}.

The generators (\ref{gen}) in the  basis (\ref{basis}) may be written
\be
T^{(n)}_\alpha := \int d^2x \ (\delta^{(n)}_\alpha a_\mu )
\ {\delta \over \delta a_\mu}
=  \int d^2x \  (D_\mu \Lambda^{(n)}_\alpha )
\ {\delta \over \delta a_\mu},
\label{basgen}
\ee
and the task is to compute the commutators
\begin{eqnarray}
[ \ T^{(m)}_\alpha, T^{(n)}_\beta\ ] &=&
\int d^2x\ \int d^2x'\  \bigl[\
\delta^{(m)}_\alpha a_\mu(x)\ {\delta \over \delta a_\mu(x)}\
D_\nu \Lambda^{(n)}_\beta(x')\  {\delta \over \delta a_\nu(x')}
\nonumber\\
& &
- \delta^{(n)}_\beta a_\nu(x')\  {\delta \over \delta a_\nu(x')}
\ D_\mu \Lambda^{(m)}_\alpha(x)\  {\delta \over \delta a_\mu(x)}
\ \bigr]
\nonumber \\
&=&  \int d^2x\   D_\mu\ \bigl[\
\delta^{(m)}_\alpha \Lambda^{(n)}_\beta
-\delta^{(n)}_\beta \Lambda^{(m)}_\alpha
+ \{ \Lambda^{(m)}_\alpha,  \Lambda^{(n)}_\beta \}\ \bigr]
\ {\delta \over \delta a_\mu}.  \label{comm}
\end{eqnarray}
In the last equality we have assumed that the surface terms arising
from integration  by parts vanish. In the following we will show that
the algebra of the $T^{(n)}_\alpha$ is the affine algebra
$[\ T^{(m)}_\alpha,\ T^{(n)}_\beta \ ] =
C_{\alpha\beta}^{\ \ \gamma}\ T^{(m+n)}_\gamma $ for $m,n\ge 0$.

We first note that  $J_\mu^{(0)} : = \{ a_\mu, \Lambda^{(0)}\}$ is a
trivial  current since its conservation is the equation of motion
$\eta^{\mu\nu}\partial_\mu a_\nu = 0$.  Taking
$ \Lambda^{(0)}_\alpha(y,z) := f_\alpha(y,z)$, and using
  (\ref{comm})  and
$\delta^{(n)}_\alpha  \Lambda^{(0)}_\beta = 0$, (which
follows because $\Lambda^{(0)}_\beta$ is independent of  $a_\mu$),
gives
\be
[ \ T^{(0)}_\alpha, T^{(0)}_\beta\ ] =
C_{\alpha\beta}^{\ \ \gamma}\ T^{(0)}_\gamma.
\ee
  Assuming  now that
\be
[ \ T^{(0)}_\alpha, T^{(n)}_\beta \ ] =
C_{\alpha\beta}^{\ \ \gamma}\ T^{(n)}_\gamma
\ee
implies
\be
D_\mu(\delta^{(0)}_\alpha  \Lambda^{(n)}_\beta ) =
C_{\alpha\beta}^{\ \ \gamma}\  D_\mu\Lambda^{(n)}_\gamma
+ \{ D_\mu \Lambda^{(n)}_\beta,  \Lambda^{(0)}_\alpha \}
-\{ \{ a_\mu,  \Lambda^{(0)}_\alpha \},  \Lambda^{(n)}_\beta \}.
\ee
Using the last equation and (\ref{lamrec}) then gives
\be
[\  T^{(0)}_\alpha, T^{(n+1)}_\beta\ ] = C_{\alpha\beta}^{\ \ \gamma}
\ T^{(n+1)}_\gamma. \label{0n+1}
\ee

The calculation of  $[\  T^{(1)}_\alpha, T^{(1)}_\beta\ ]$ requires
$\delta^{(1)}_\alpha \Lambda^{(1)}_\beta$. In general for $n>0$ we
have  using (\ref{amu}), (\ref{lam1}) and (\ref{lamrec}) that
\be
\delta^{(n)}_\alpha \Lambda^{(1)}_\beta =
\{ \delta^{(n)}_\alpha \phi,  \Lambda^{(0)}_\beta \} =
 \{ \int _{-\infty}^x dx'
D_0  \Lambda^{(n)}_\alpha,  \Lambda^{(0)}_\beta \}
=  \{  \Lambda^{(n+1)}_\alpha,  \Lambda^{(0)}_\beta \},
\ee
 Assuming   that
\be
[\  T^{(1)}_\alpha, T^{(n)}_\beta \ ] =
C_{\alpha\beta}^\gamma\ T^{(n+1)}_{\ \ \gamma}
\label{1n}\ee
implies
\be
 \delta^{(1)}_\alpha \ \Lambda^{(n)}_\beta  =
C_{\alpha\beta}^{\ \ \gamma}\  \Lambda^{(n+1)}_\gamma
+ \{   \Lambda^{(n+1)}_\beta,  \Lambda^{(0)}_\alpha \}
-\{   \Lambda^{(1)}_\alpha ,  \Lambda^{(n)}_\beta \}.
\ee
Using this formula and (\ref{lamrec}) gives
\begin{eqnarray}
 \delta^{(1)}_\alpha\  \Lambda^{(n+1)}_\beta &=&
\int_{-\infty}^x dx\  \bigl[\  D_0\  (\delta^{(1)}_\alpha
 \Lambda^{(n)}_\beta)
+ \{ D_0   \Lambda^{(1)}_\alpha,  \Lambda^{(n)}_\beta \}\  \bigr]
\nonumber \\
&=& \int_{-\infty}^x dx \ \bigl[ \
C_{\alpha\beta}^{\ \ \gamma}\  D_0\Lambda^{(n+1)}_\gamma
+ D_0\{   \Lambda^{(n+1)}_\beta,  \Lambda^{(0)}_\alpha \}
-\{   \Lambda^{(1)}_\alpha , D_0 \Lambda^{(n)}_\beta \}\ \bigr]
\nonumber \\
&=& C_{\alpha\beta}^{\ \ \gamma}\  \Lambda^{(n+2)}_\gamma
+  \{   \Lambda^{(n+2)}_\beta,  \Lambda^{(0)}_\alpha \}
- \int_{-\infty}^x dx \ \bigl(\
\{   \Lambda^{(1)}_\alpha,  D_0\Lambda^{(n)}_\beta \}
- \{   \Lambda^{(n+1)}_\beta,  \{ a_0, \Lambda^{(0)}_\alpha \} \}
\ \bigr)
\nonumber \\
&=& C_{\alpha\beta}^{\ \ \gamma}  \Lambda^{(n+2)}_\gamma
+ \{   \Lambda^{(n+2)}_\beta,  \Lambda^{(0)}_\alpha \}
-\{   \Lambda^{(1)}_\alpha ,  \Lambda^{(n+1)}_\beta \}.
\end{eqnarray}
It  therefore follows from (\ref{comm}) that
\be
[ \ T^{(1)}_\alpha,\ T^{(n+1)}_\beta\ ] =
C_{\alpha\beta}^{\ \ \gamma}\ T^{(n+2)}_\gamma, \label{1n+1}
\ee
which completes the induction.  It remains to show that
\be
[\  T^{(m)}_\alpha,\ T^{(n)}_\beta\ ] =
C_{\alpha\beta}^{\ \ \gamma}\ T^{(m+n)}_\gamma,
\label{gencom}
\ee
for all $m,n$. This  can again be proved by induction using (\ref{1n})
and the Jacobi identity.  Assuming (\ref{gencom}), we have
\begin{eqnarray}
 C_{\sigma\rho}^{\ \ \beta}\ [\  T^{(n)}_\alpha,\ T^{(m+1)}_\beta\  ] &=&
  [\  T^{(n)}_\alpha,\ [\   T^{(m)}_\sigma,\ T^{(1)}_\rho\   ] \  ]
\nonumber \\
&=& -  [\  T^{(m)}_\sigma,\  [\  T^{(1)}_\rho,T^{(n)}_\alpha \  ] \  ]
 -  [\  T^{(1)}_\rho,\  [\  T^{(n)}_\alpha,\ T^{(m)}_\sigma  \  ]\ ]
   \nonumber \\
&=& -C_ {\rho\alpha}^{\ \ \gamma} \
   [\ T^{(m)}_\sigma,\ T^{(n+1)}_\gamma\ ]
+ C_{\sigma\alpha }^{\ \ \gamma}\ [\ T^{(1)}_\rho ,\ T^{(m+n)}_\gamma\ ]
\nonumber \\
&=& - C_ {\rho\alpha}^{\ \ \gamma}\ [\ T^{(m)}_\sigma,\ T^{(n+1)}_\gamma\ ]
 + C_{\sigma\alpha}^{\ \ \gamma}\ C_{\rho\gamma}^{\ \ \beta}\
T^{(m+n+1)}_\beta.
\end{eqnarray}
Therefore
\be
C_{\sigma\rho}^{\ \ \beta}\ [\  T^{(n)}_\alpha,\ T^{(m+1)}_\beta\  ]
+ C_ {\rho\alpha}^{\ \ \beta}\  [\ T^{(m)}_\sigma,\ T^{(n+1)}_\beta \ ]
= C_{\sigma\alpha}^{\ \ \beta}\ C_{\rho\beta}^{\ \ \gamma}\
T^{(m+n+1)}_\gamma.
\label{expan}
\ee
Expanding the second term on the r.h.s. gives
\begin{eqnarray}
C_ {\rho\alpha}^{\ \ \beta}\  [\ T^{(m)}_\sigma,\ T^{(n+1)}_\beta \ ]
&=& -[\ T_\rho^{(n)},\ [\ T_\alpha^{(1)},\ T_\sigma^{(m)}\ ] \ ]
- [\ T_\alpha^{(1)},\ [\  T_\sigma^{(m)},\ T_\rho^{(n)} \ ] \ ]
\nonumber \\
&=& -C_{\alpha\sigma}^{\ \ \beta}\ [\ T_\rho^{(n)},\  T^{(m+1)}_\beta \ ]
 - C_{\sigma\rho}^{\ \ \beta}\ C_{\alpha\beta}^{\ \ \gamma}\
T^{(m+n+1)}_\gamma.
\end{eqnarray}
Substituting this back into (\ref{expan}) gives
\be
C_{\sigma\rho}^{\ \ \beta}\ [\  T^{(n)}_\alpha,\ T^{(m+1)}_\beta\  ]
+ C_{\sigma\alpha}^{\ \ \beta}\ [\ T_\rho^{(n)},\  T^{(m+1)}_\beta \ ]
= (\ C_{\sigma\rho}^{\ \ \beta}\ C_{\alpha\beta}^{\ \ \gamma}
+ C_{\sigma\alpha}^{\ \ \beta}\ C_{\rho\beta}^{\ \ \gamma}\ )
\ T^{(m+n+1)}_\gamma.
\ee
This completes the proof that the hidden symmetries of  self-dual
gravity form the Affine algebra associated with the Lie algebra
of area preserving diffeomorphisms.

We note that for $m=n=0$, the algebra (\ref{gencom}) is that of area
preserving diffeomorphisms, or $w_\infty$, which is the special case
identified previously \cite{park}. We note also that there is
an infinite dimensional commuting set of generators obtained
by setting the Lie algebra indices in (\ref{gencom}) to be equal:
$\alpha=\beta$.
This commuting set  may provide the proof of integrability in the
Liouville sense, if it can be shown that it also arises from two
distinct Hamiltonian formulations in the manner standard for
two dimensional integrable models \cite{das}.

The symmetry transformations that we have found also provide a
solution generating technique, whereby given one solution, there
is a method for obtaining a new solution.  This is analogous to
the procedure given by Geroch for two Killing field reductions
of general relativity \cite{ger}.

We consider a one parameter set of solutions $a_\mu(s; t,x;y,z)$
such that $s=0$  is the given solution. Then a new solution, specified
by the parameter value $s=1$, is obtained by integrating the
equation
\be
{d a_\mu\over ds} = F^\alpha_n(s)\ \delta^{(n)}_\alpha a_\mu =
F^\alpha_n(s)\ D_\mu \Lambda^{(n)}_\alpha,
\label{solugen}
\ee
where $F^\alpha_n(s)$ are arbitrary functions which will characterize
the new solution.  (There is an implied sum over repeated indices).

This procedure for generating solutions is in fact general, and applies
to any theory  where hidden (non-gauge) symmetries exist on the
solution space.  If we consider the Hamiltonian formulation
of the self-dual gravity chiral model, the  phase space analog of
(\ref{solugen}) would involve  writing the hidden symmetry generators
as  functionals of the phase space variables, and replacing
$ \delta^{(n)}_\alpha a_\mu $ in (\ref{solugen}) by the Poisson
bracket.

\section{Discussion}

We have studied  self-dual gravity in its formulation as the two
dimensional  principal chiral model,  and obtained a result at the
classical level which identifies the observable algebra for this
theory.

There are  two main  questions that remain for the classical theory:
(1) Can the Hamiltonian formulation of the $SU(N)$ chiral model be
generalized to the infinite dimensional Lie algebra case relevant
for self-dual gravity, and (2) can the phase space analogs of
the Affine symmetry generators (\ref{basgen}) be determined
in the Hamiltonian formulation of this chiral model? Here the
Affine algebra should arise as the Poisson algebra of its
phase space generators.

For addressing the quantization problem, one is led to the problem of
finding representations of the Affine algebra associated with the Lie
algebra of area preserving diffeomorphisms. The only representations
that have been studied are those for  $SU(N)$ \cite{goddol}, so this
appears to be an open problem.

In the canonical approach to the quantization of  general relativity
motivated by the Ashtekar variables \cite{ash},  knot theory
enters in the characterization of physical states \cite{rs}.  This is
due  essentially to the imposition of the spatial diffeomorphism
constraint on functions of loops.  Because of this,  any theory
that contains a three dimensional spatial  diffeomorphism constraint
will have physical states labelled by knot classes. An example of
such a theory is  given in Ref.  \cite{kv}.

Since self-dual gravity is a diffeomorphism invariant theory, one
can ask how the connection with  knot theory arises for this case.
In the chiral model formulation discussed here, we have made some
coordinate choices and hence fixed some of the diffeomorphism
freedom.  Therefore the connection with knot theory cannot arise in
the same way as in \cite{rs}. Nevertheless it seems  possible that
 a connection  occurs via a different route. In the standard approach
to  the quantum integrability of two dimensional models, the Yang-Baxter
equation arises. This equation is connected with knot theory in that
it contains the Reidmeister moves, and furthermore, it provided the
first instances of  quantum group structures in physical problems
\cite{kauf}.   Since the $SU(N)$ chiral model is integrable
classically and quantum mechanically,  it is possible that the
chiral model approach to self-dual gravity will provide  connections
to quantum groups and knot theory in a similar way. The main issue
that needs to be addressed  for this is to see how, if at all,
the presence of an infinite dimensional Lie algebra affects the
integrability procedures for two dimensional models.

\bigskip
This work was supported by the Natural Science and Engineering
Research Council of Canada.

 \end{document}